\documentclass[aps,prx,floatfix,twocolumn,superscriptaddress,10pt]{revtex4-2}
\usepackage{amsmath,amssymb,amsthm}
\usepackage{graphicx}
\usepackage{color}
\usepackage{hyperref}
\hypersetup{
    colorlinks=true,
    linkcolor=blue,
    filecolor=magenta,      
    urlcolor=blue,
    citecolor=blue,
    pdfpagemode=FullScreen,
}

\begin{document}

\newcommand{\Uf}{U_\text{f}}
\newcommand{\uf}{u_\text{f}}
\newcommand{\aSA}{\alpha_\text{\tiny SA}}
\newcommand{\aFMS}{\alpha_\text{\tiny FMS}}
\newcommand{\acr}{\alpha_\text{cross}}
\newcommand{\aalg}{\alpha_\text{alg}}
\newcommand{\aZTMC}{\alpha_\text{\tiny G}}
\newcommand{\Psat}{P_\text{\tiny SAT}}

\title{Algorithmic thresholds in combinatorial optimization depend on the time scaling}

\author{M. C. Angelini}
\affiliation{Dipartimento di Fisica, Sapienza Università di Roma, P.le Aldo Moro 5, 00185 Rome, Italy}
\affiliation{Istituto Nazionale di Fisica Nucleare, Sezione di Roma I, P.le A. Moro 5, 00185 Rome, Italy}

\author{M. Avila-Gonz\'alez}
\affiliation{Group of Complex Systems and Statistical Physics and  Department of Theoretical Physics. Physics Faculty, University of Havana. CP10400, La Habana, Cuba}

\author{F. D'Amico}
\affiliation{Dipartimento di Fisica, Sapienza Università di Roma, P.le Aldo Moro 5, 00185 Rome, Italy}
\affiliation{CNR - Nanotec, unit\`a di Roma, P.le Aldo Moro 5, 00185 Rome, Italy}

\author{D. Machado}
\affiliation{Dipartimento di Fisica, Sapienza Università di Roma, P.le Aldo Moro 5, 00185 Rome, Italy}
\affiliation{Group of Complex Systems and Statistical Physics and  Department of Theoretical Physics. Physics Faculty, University of Havana. CP10400, La Habana, Cuba}
\affiliation{CNR - Nanotec, unit\`a di Roma, P.le Aldo Moro 5, 00185 Rome, Italy}

\author{R. Mulet}
\affiliation{Group of Complex Systems and Statistical Physics and  Department of Theoretical Physics. Physics Faculty, University of Havana. CP10400, La Habana, Cuba}

\author{F. Ricci-Tersenghi}
\affiliation{Dipartimento di Fisica, Sapienza Università di Roma, P.le Aldo Moro 5, 00185 Rome, Italy}
\affiliation{Istituto Nazionale di Fisica Nucleare, Sezione di Roma I, P.le A. Moro 5, 00185 Rome, Italy}
\affiliation{CNR - Nanotec, unit\`a di Roma, P.le Aldo Moro 5, 00185 Rome, Italy}

\begin{abstract}
In the last decades, many efforts have focused on analyzing typical-case hardness in optimization and inference problems. Some recent work has pointed out that polynomial algorithms exist, running with a time that grows more than linearly with the system size, which can do better than linear algorithms, finding solutions to random problems in a wider range of parameters. However, a theory for polynomial and superlinear algorithms is in general lacking.
In this paper, we examine the performance of the Simulated Annealing algorithm, a standard, versatile, and robust choice for solving optimization and inference problems, in the prototypical random $K$-Sat problem. For the first time, we show that the algorithmic thresholds depend on the time scaling of the algorithm with the size of the system. Indeed, one can identify not just one, but different thresholds for linear, quadratic, cubic regimes (and so on). This observation opens new directions in studying the typical case hardness in optimization problems.
\end{abstract}

\maketitle

\section{Introduction} 

Computing the performance of algorithms solving optimization and inference problems is crucial for both, fundamental research and real-world applications. In particular, it is primarily important to identify the so-called \emph{algorithmic threshold}, that is the limit of the region where a specific algorithm can work efficiently to find a solution to the problem.

In many relevant problems, the best known algorithmic threshold is far from the \emph{statistical threshold}, which can be obtained by computing the limit of the existence of solutions. In other words, there exists a region (often very broad) where solutions are likely to be present, but no algorithm can find them efficiently. This region is called the \emph{hard phase} and it is at the origin of the computational-to-statistical gap that plagues many optimization and inference problems.
To close, or at least reduce, this gap one should move forward the algorithmic threshold.

The vast majority of algorithms whose performances can be analyzed are linear algorithms, that is, algorithms whose running time scales linearly with the problem size. This is a major limitation in several aspects. First of all, superlinear but still polynomial algorithms may show better performances and may have improved thresholds. Not every algorithm can be scaled up to a superlinear regime, but even the scalable ones have not been studied so far. The present work represents the first systematic study of the superlinear regime in a general class of algorithms. We expect that it may open a new approach to the determination of algorithmic thresholds.

Secondly, the algorithmic threshold for linear algorithms is often computed via extrapolations from the linear regime as the point where the time to reach a solution diverges (see e.g.\ insets in Figs.~1 and 5 of Ref.~\cite{AlavaPNAS2008}). These estimates not only require taking the large size and large time limits, but are unavoidably quite noisy due to the extrapolation procedure. Here we show how to evaluate algorithmic thresholds in a much better way, by going to the superlinear regime (in particular the quadratic regime) and exploiting well-known techniques from the study of critical phenomena. We consider that this new approach can become the standard for the determination of algorithmic thresholds.

To show how to achieve the goals stated above we choose to focus our study on a problem and an algorithm that are general, representative, and with non-trivial properties. In particular, we choose the random $K$-Satisfiability problem and the Simulated Annealing algorithm. Both have widespread applications in many diverse fields of science and technology.

The $K$-Satisfiability ($K$-Sat) problem is a pillar in the theory of algorithmic complexity. It was the first to be classified as NP-complete, actually, in the same article where Cook introduced the very concept of NP-completeness \cite{cook1971KSAT}. The task in $K$-Sat is to find an assignment to $N$ boolean variables $\vec{x} = \{x_1, \ldots, x_N \}$ that satisfies a set of $M$ clauses that involve $K$ variables each. In the random $K$-Sat problem, the $K$ variables in each clause are chosen at random, and thus the hardness of the formula is controlled mainly by the parameter $\alpha = M / N$. This problem undergoes a Sat-Unsat transition at a satisfiability threshold $\alpha_s$ \cite{mezardksat2002}, such that for $\alpha < \alpha_s$ solutions exist almost surely in the infinite size limit ($N \to \infty$) while for $\alpha > \alpha_s$ in the same limit there are no solutions with high probability: this transition has been proved to be a sharp one in the large size limit \cite{friedgut1999sharp,ding2015proof}.

One could associate with each configuration of the $N$ variables a cost function that counts the number of violated clauses. In this way, the problem of finding a satisfying assignment is recast as an optimization problem, for which one wants to identify a configuration of zero cost.
The $K$-Sat problem is thus seen as the paradigm of combinatorial optimization problems, for this reason, we will concentrate our attention on this problem, keeping in mind that what is discussed in the following is shared by many other random optimization problems, as we discuss at the end of the paper, presenting results for a different random optimization problem.

While the ``worst-case'' NP-completeness of $K$-Sat problems has been stated many years ago \cite{cook1971KSAT}, in more recent years many efforts have been focused on studying typical-case hardness \cite{mezard2002random, mezard2005clustering, MertensKSAT2006, achlioptas2006solution, KrzakalaKSAT2007,gamarnik2018finding}.
One would like to identify the best algorithm that can find optimal configurations (i.e., solutions) of the problem within the shortest time in typical instances of the problem, for which we know that solutions exist up to $\alpha_s$ in the large size limit. However, in practice,  search algorithms are not able to find solutions, although they do exist, in a region $\alpha_\text{alg}<\alpha<\alpha_s$ \cite{coja2010better}. The precise determination of the threshold $\alpha_\text{alg}$ in typical instances, which could be specific to every algorithm, has been the object of an intense effort in the last decades \cite{Seitz_2005,AlavaPNAS2008, marino_backtracking_2016, Budzynski_2019, angelini2019monte, angelini2023limits} because for practical applications, the algorithmic threshold $\alpha_\text{alg}$ matters much more than the Sat-Unsat transition at $\alpha_s$.

Exploiting the so-called one-step Replica Symmetry Breaking (1-RSB) Cavity Method formalism, theoretical physicists have characterized different phase transitions in the geometrical structure of the space of solutions \cite{MertensKSAT2006, KrzakalaKSAT2007, Montanari_2008}. This structure is certainly related to the algorithmic complexity of the problem, but a clear connection is still missing. In the following, we summarize the phase transitions taking place in random $K$-Sat and several conjectures relating phase transitions to algorithm thresholds.

Increasing $\alpha$, the first phase transition the model undergoes is the \emph{dynamical transition} at $\alpha_d$. Above $\alpha_d$, the uniform equilibrium measure over the solutions breaks into an exponential number of disconnected clusters \cite{mezard2005clustering, achlioptas2006solution, KrzakalaKSAT2007}. The \emph{condensation transition} takes place at $\alpha_c \geq \alpha_d$ and defines the threshold above which the equilibrium measure is dominated by a sub-exponential number of clusters \cite{KrzakalaKSAT2007}. Given that above $\alpha_d$ the ergodicity is spontaneously broken (because of the existence of many clusters separated by barriers), it was initially conjectured $\alpha_d$ to be an upper bound for the local search algorithms \cite{mezardksat2002}.
Later, it was found that the Belief Propagation algorithm returns exact marginals until $\alpha_c$, and consequently, $\alpha_c$ was conjectured to be the algorithmic threshold \cite{KrzakalaKSAT2007}. This conjecture was further improved in \cite{ricci2009cavity}.

However, there are several observations of local search algorithms overcoming both bounds $\alpha_d$ and $\alpha_c$ \cite{Seitz_2005, AlavaPNAS2008, angelini2019monte} solving almost all typical instances in a time that scales polynomially with the problem size for $\alpha$ very close to the satisfiability threshold $\alpha_s$ in some optimization problems (included the $K$-Sat problem for small enough $K$). An intuitive solution to this contradiction \cite{LemoyVFMS2015} rests on the idea that some algorithms violate the detailed balance condition and are completely out of equilibrium. Therefore, they do not necessarily fit within the hypothesis of the equilibrium computations that lead to the theoretical thresholds $\alpha_d$ and $\alpha_c$. 
Even speaking about algorithms satisfying the detailed balance condition, like Monte Carlo algorithms, they are often not used in the regime of very large times, but on a finite time scale: in such a regime, the algorithm could be visiting configurations with a measure different from the equilibrium one. Their performances are thus not well described by equilibrium computations.

Forgetting about the equilibrium measure, one could ``follow'' the states out of equilibrium \cite{zdeborova2010generalization}. By doing so, even above the clustering threshold, there could exist a phase displaying the so-called ``canyon-dominated landscape'' where out-of-equilibrium states go down to zero energy with a large basin of attraction. The region of the parameters where these canyon states do not exist anymore could correspond to the hard phase for algorithms. This hypothesis was checked in p-spin glass models for the Simulated Annealing algorithm \cite{krzakala2013performance}. The canyon-dominated threshold is upper-bounded by the so-called \emph{freezing transition}, above which any cluster of solutions contains an extensive number of variables that are frozen, that is, taking the same value in all the solutions belonging to the cluster \cite{achlioptas2006solution, ardelius2008exhaustive}. Finding solutions in frozen clusters is very hard as no error can be made for any frozen variable. Being the number of frozen variables extensive, the probability of making no error fixing them goes to zero exponentially fast in the problem size. For this reason, the freezing transition has been conjectured to be an upper bound to the algorithmic threshold (again, under the hypothesis that solutions are sampled according to the equilibrium measure).

A new approach to algorithmic hardness in random problems was recently introduced: the so-called Overlap Gap Property (OGP), based on the topology of the solution space \cite{gamarnik2021overlap}. Such a property, already discovered in Refs.~\cite{achlioptas2006solution,mezard2005clustering} but named OGP only in Ref.~\cite{gamarnik2018finding}, allows one to rigorously demonstrate the ineffectiveness of a large class of algorithms, named \emph{stable algorithms} \cite{gamarnik2017limits}, in solving the corresponding optimization problem. Stable algorithms include low-degree polynomial algorithms \cite{hopkins2017efficient, hopkins2017power, hopkins2018statistical}, approximate message passing algorithms \cite{zdeborova2016statistical} and Monte Carlo algorithms run for a fixed (i.e.\ not growing with $N$) number of steps. One could conjecture the OGP to be plausible evidence for computational intractability, not just for stable algorithms, but even for more general algorithms.
There are, however, counterexamples: in a recent work \cite{li2024some}, it has been proved that the shortest-path optimization problem has the OGP in sparse graphs, but it can be solved by $\mathcal{O}(\log N)$-degree polynomial algorithms. Thus, in this case, the OGP is not predictive of algorithmic intractability for an average-case optimization problem.
Something similar also happens in the case of the XOR-Sat problem \cite{RicciPspin2001}, which displays the OGP \cite{mezard2003two} while it is easy to solve through Gaussian elimination algorithms \cite{ricci2010being}.
Both the exact nature of the stability of the algorithms and the existence of the OGP are problem-dependent and sometimes extremely technical to prove.

In this paper, we will analyze the performances of algorithms based on Monte Carlo Markov Chains (MC in short), in particular, Simulated Annealing (SA) \cite{KirkpatrickSA1983} and the zero-temperature Monte Carlo (greedy algorithm), both well-known in statistical physics. For the interested reader, we make all the code and data necessary to reproduce our results available online \cite{tScalingAlgThresholds}. Despite their robustness and flexibility, properties that make them reference algorithms in many situations (especially when real-life instead of random problems are considered), there are no known works that precisely identify the thresholds reached by these algorithms on the prototypical random $K$-Sat optimization problem.
First, we will show that, even if detailed balance is met, these algorithms can run very efficiently in $K$-Sat instances, finding solutions beyond the dynamical threshold $\alpha_d$ and even beyond the condensation threshold $\alpha_c$ for small $K$. 

We will then introduce a new crucial ingredient in the analysis of these algorithms: the scaling of the running time with the size of the system. 
A Monte Carlo Sweep (MCS) is defined as the attempt to change every variable in the configuration once.
MC-based algorithms are local algorithms when they are run for a number of MCS that is fixed with respect to the size of the problem. In this way, the running time of the algorithm scales linearly with the problem size $N$, because the single MCS requires a time $t=\mathcal{O}(N)$. Under these conditions, one can often demonstrate that MC algorithms are stable algorithms, and their performances are thus limited by the existence of OGP.
However, when running for a number of MCS that increases with the problem size, MC algorithms are not stable algorithms anymore, and nothing is known about their performance.

In recent literature, there have been few works showing that polynomial algorithms, with running time that scales at least quadratically with the system size, could beat very efficient linear time algorithms in different optimization and inference problems \cite{angelini2018parallel, song2021cryptographic, zadik2022lattice, Erba_2024}.
In this paper, for the first time, we identify different thresholds for MC algorithms, depending on the scaling of the number of MCS with the problem size $N$, i.e., linear, quadratic, cubic, and so on.
MC algorithms are ideal for running such an experiment, given that one can decide the total number of MCS before running the algorithm, and such a number can be varied when the system size increases.

We expect the behavior we uncover in this work to be possibly shared by other classes of algorithms. For example, neural networks can be trained and tested for a number of epochs that could be changed accordingly with the size of the problem. 
Our work thus opens new perspectives on the study of typical case hardness: how could we explain theoretically the presence of different thresholds related to different polynomial time scaling?

\section{Methods}

The standard methodology to determine algorithmic thresholds measures the time an algorithm needs to find a solution (in short, time-to-solution or $TTS$) and studies how it scales with the system size at different values of $\alpha$. However, this analysis does not provide a clear estimate for the algorithmic threshold. Indeed, for the sizes one can study, $TTS(N)$ makes just a smooth crossover from a ``slow'' to a ``fast'' growth around the algorithmic threshold.

We follow, instead, two other routes for the analysis of the performance of MC-based algorithms in solving $K$-Sat problems and the determination of the algorithmic threshold for SA. Our two methods give consistent results for the $K$-Sat problem with $K=3,4$.

First, we use an approach that exploits finite-size effects to determine the algorithmic threshold $\aSA$, similar to the approach used to study the critical behavior in statistical physics models. We run the algorithm (e.g., SA in our case) for a polynomial time $t \sim N^z$, where the ``dynamical'' exponent $z$ controls the time scaling.
Then we examine the mean extensive energy $U$ reached after a time $t$ or the probability of solving the problem $\Psat$ in a time $t$, as a function of $\alpha$. Both quantities, $U$ and $\Psat$, have a clear crossing at the same algorithmic threshold $\aSA$, supporting the scenario that below $\aSA$ every problem can be solved in the large $N$ limit if SA is run for a time $t=O(N^z)$, while none is solvable by SA above $\aSA$.
We will show how this algorithmic threshold depends on the value of $z$, that is on the time scaling chosen for running the algorithm.

Our main results are obtained for $z=2,3$ (i.e., quadratic and cubic running times).
These results provide clear evidence that the algorithmic threshold depends on the time scaling.

Determining the SA threshold for times growing less than quadratic (e.g., linear times) is a more difficult task.
A MCS corresponds to the update of every variable once, and it takes a time $O(N)$.
Thus, the linear time regime actually corresponds to running a MC algorithm for a \emph{fixed} (not growing with $N$) number of MCS.
Even below the algorithmic threshold, it is not surprising that running an MC algorithm for a fixed number of MCS makes the success probability dependent on the system size (the number of MCS that are enough for a given problem are probably not enough for a much larger problem).
Indeed, in this linear-time regime, we find severe finite-size effects that make the techniques we have used for $z=2,3$ useless.
To overcome this problem, we have studied MC at zero temperature, which corresponds to a sort of greedy algorithm without thermal noise, that we find to run in almost linear time.
Results are reported in Appendix~\ref{app:linearRegime}.

As a complementary approach, we also work in the ``infinite-size limit'', where this limit must be understood as ``the system size is large enough that finite-size effects are negligible''. In this limit, self-averaging quantities have vanishing fluctuations and concentrate on a unique curve (and from this unique curve, we need to extract the requested information). For example, considering the intensive energy $u$ reached after a running time $t$, one expects a ``relatively fast'' decay of $u(t)$ below the algorithmic threshold and a ``much slower'' decay above the threshold.
We will provide more quantitative statements below, but in practice, in this infinite-size limit, the threshold value $\aSA^{\infty}$ is defined as the critical value such that $u(t)$ decays as a power law of the running time $t$.
This is very similar to what has been observed in Ref.~\cite{Budzynski_2019} for the hypergraph bi-coloring problem.
It is worth stressing that in this infinite-size limit, the time required to solve the problem is always formally infinite, and this is the reason why the algorithmic threshold must be estimated from the way the energy decays for large times, rather than from the time to reach a solution.



\section{The $K$-Sat problem and the Simulated annealing algorithm} \label{sec:SA}

The task in $K$-Sat is to find a satisfying assignment for $N$ boolean variables $\vec{x} = \{x_1, \ldots, x_N \}$ that participate in a Boolean expression $f(\vec{x})$. The hardness of the problem is in the structure of the formula $f$, which is the conjunction (logic \textit{and}) of $M$ smaller expressions called \emph{clauses}. A clause is a disjunction (logic \textit{or}) of $K$ variables or their negations. The following simple example has $K=3$, $M=2$ clauses and only $N=3$ variables: 
\begin{equation*}
 f(x_1, x_2, x_3) = (x_1 \lor x_2 \lor x_3) \land (-x_1 \lor -x_2 \lor x_3).
\end{equation*}
While in this example it is easy to find a satisfying assignment by simple inspection, this is not necessarily the case when $N$ and $M$ increase. In the random $K$-Sat, the variables entering each clause are randomly chosen among the $N$ possible ones and negated with probability 0.5. The parameter $\alpha = M / N$ controls the hardness of the formulas. Empirical evidence and theoretical studies \cite{mezard2005clustering, MertensKSAT2006, KrzakalaKSAT2007,Montanari_2008,Braunstein_2016,gamarnik2018finding} support the existence of a hard phase before the Sat-Unsat threshold: in the hard phase, solutions do exist, but searching algorithms require times scaling super-polynomially with the problem size to find a solution.

In combinatorial optimization problems, the most natural choice for the energy function is the number of unsatisfied clauses $U(\vec{x})$. This energy function is bounded between 0 and $M$, reaching the minimum value $U=0$ for every satisfying assignment. 

MC algorithms are the standard technique for sampling the configuration space of the system according to the Boltzmann-Gibbs distribution $P(\vec{x}) = e^{-\beta U(\vec{x})} / Z$. Here, $Z$ is a normalization factor and $\beta=1/T$ is the inverse temperature. While solutions of the optimization problem correspond to the equilibrium configurations at $T=0$, a positive temperature allows \emph{uphill} moves that facilitate overcoming the local barriers.

A single MC step implementing the Metropolis dynamics \cite{Metropolis1953} works as follows. Given a configuration $\vec{x}$, a new configuration $\vec{x}'$ is proposed where a single variable is flipped. The variable to be flipped is chosen uniformly at random among the $N$ possible ones. The proposed change in configuration is accepted with probability
\begin{equation}
r(\vec{x} \to \vec{x}') =
\left\{
\begin{array}{ll}
1 & \text{if} \:\: \Delta U(\vec{x} \to \vec{x}') \leq 0 \\
e^{-\beta\,\Delta U(\vec{x} \to \vec{x}')} & \text{if} \:\: \Delta U(\vec{x} \to \vec{x}') > 0
\end{array}
\right.
\label{eq:Metropolis_rates}
\end{equation}
where $\Delta U(\vec{x} \to \vec{x}') = U(\vec{x}') - U(\vec{x})$.
The time unit of MC algorithms is the MCS, consisting of $N$ MC steps. That is, in a MCS every variable of the problem is updated on average once.

The SA algorithm starts implementing MC at a high temperature (low $\beta$), where there are no barriers and the configuration can be easily updated. Then the temperature $T$ is slowly decreased, performing some MCS at each selected temperature. This can be implemented using different schemes \cite{EgleseSimAnn1990, GuilmeauSimAnn2021}. For simplicity, we restricted ourselves to the traditional choice of a constant temperature change $\Delta T$ \cite{KirkpatrickSA1983}. 
The aim is to have the SA algorithm that keeps sampling equilibrium configurations while decreasing the temperature. If this can be achieved until $T=0$, then a solution with $U=0$ can be obtained (if it exists). In practice, when dealing with finite size systems and finite running times, we do not expect SA to be able to sample from the equilibrium distribution at each time. From this point of view, real-world implementations of SA are typically out-of-equilibrium processes.

The \emph{cooling schedule} is the sequence of chosen temperatures $(T_0, T_1, \ldots, T_n)$ used by SA and the number of MCS the algorithm will take at each temperature. In general, the initial temperature $T_0$ is required to be high enough to be in the ergodic phase and the final temperature $T_n$ should be (close to) $T=0$. 
In our simulations, we consider a very simple schedule (from $T_0$ to $T_n=0$ with fixed $\Delta T=T_0/n$), and one MCS is performed at each temperature, such that the SA algorithm is run for a total of $n$ MCS.
Although the schedule is probably not the optimal one, we choose it to be quasi-adiabatic (the $\Delta T$ value is the smallest possible) and we expect that changing $T_0$ and $T_n$ would change only prefactors and not the scaling between size and time (which is what we are interested in).
In general, it is advisable to use a simple schedule that can provide a clear indication of the time scaling, although the absolute running time could be further optimized.

The reason why it is worth studying the SA algorithm is that it is known to be general (i.e., it can be used to optimize any energy function) and very robust.
The use of the temperature is a standard approach when exploring a rough energy landscape to allow for the bypassing of energy barriers.
The slow convergence to the zero-temperature limit focuses the Gibbs measure on the solutions to the problem, thus formally ensuring the optimum could be achieved in the infinite time limit.

\begin{figure}[t]
\centering
\includegraphics[width=0.4\textwidth]{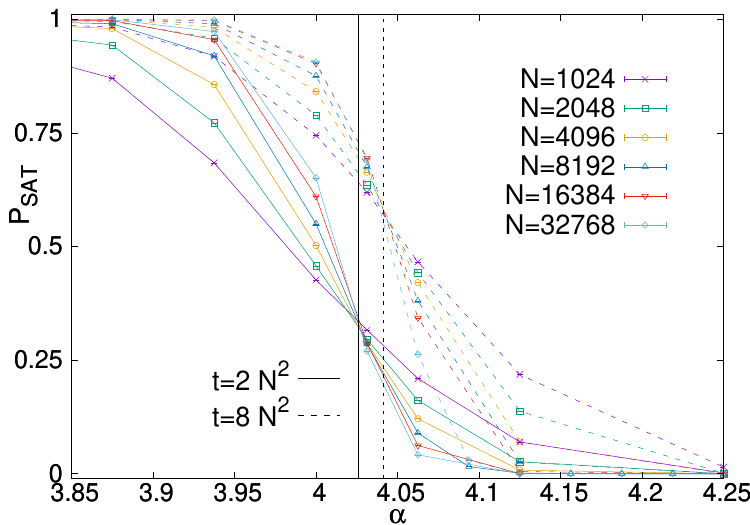}
\includegraphics[width=0.4\textwidth]{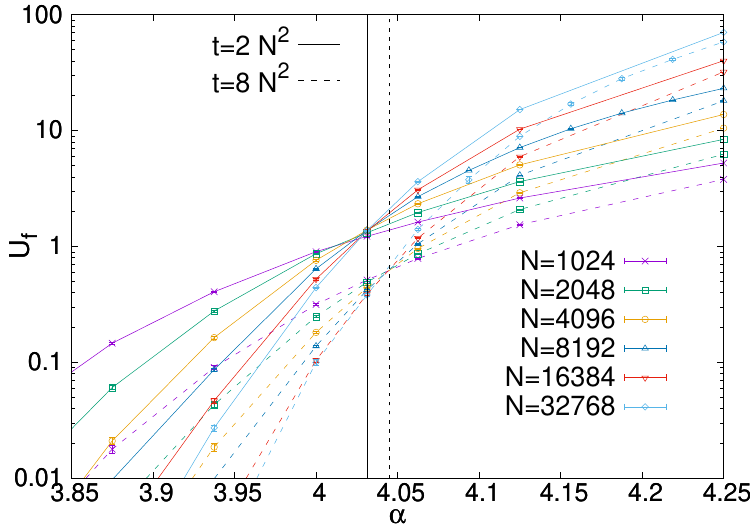}
\caption{{\bf Simulated annealing thresholds for quadratic time scaling in random 3-Sat instances}. For several system sizes, we measure the probability $\Psat$ of finding a solution (upper panel) and the average extensive energy $\Uf$ (lower panel) after the Simulated Annealing algorithm has been run for a quadratic time, $t=A N^2$. Continuous lines are for $A=2$ and dashed lines are for $A=8$. The algorithmic thresholds are $\aSA^{N^{2}} = 4.031(2)$ for $A=2$ and $\aSA^{N^{2}} = 4.045(5)$ for $A=8$.}
\label{fig:SA_transition_other_scalings}
\end{figure}

\section{Algorithmic thresholds from finite size analysis}
\label{sec:quadratic}

In many cases, the running time of algorithms based on MC, such as SA, is limited to a fixed number of MCS and, therefore, grows linearly with the system size. Unfortunately, in this linear time regime, it is not easy to estimate the algorithmic threshold. We discuss this regime for SA solving K-Sat in Appendix~\ref{app:linearRegime}. Here, we show how better algorithmic thresholds can be estimated by going to superlinear time scalings. 

To explore this idea, we ran SA with the schedule mentioned above for different system sizes $N$ and a number of MCS growing linearly in $N$, implying $t=O(N^2)$. As expected, in the upper panel of Fig.~\ref{fig:SA_transition_other_scalings}, the probability of finding a solution $\Psat$ decreases as $\alpha$ increases. The drop in the probability becomes sharper as $N$ increases. All the curves cross at the same value of $\alpha$ (within the statistical uncertainty of our data), which represents the algorithmic threshold $\aSA$. Indeed, we expect the curves to become a step function in the large $N$ limit, such that $\Psat=1$ for $\alpha<\aSA$ and $\Psat=0$ for $\alpha>\aSA$.

The bottom panel of Fig.~\ref{fig:SA_transition_other_scalings} shows an analogous behavior in the final mean extensive energy $\Uf$ achieved by SA with the same quadratic scaling. The final energy grows with $\alpha$ with a slope increasing with $N$. Data for different problem sizes crosses at the same $\aSA$ computed from the crossing of $\Psat$ curves, providing stronger and consistent evidence for a sharp algorithmic transition at $\aSA$. Indeed, in the large $N$ limit, $\Uf$ tends to zero for $\alpha<\aSA$, while it diverges for $\alpha>\aSA$.

It is worth stressing that the nice crossings in Fig.~\ref{fig:SA_transition_other_scalings} have never been observed before because MC algorithms (as many other algorithms) were studied only in the linear regime. A reliable algorithmic threshold for SA can be obtained only by scaling times at least quadratically with the problem size.

The specific value of $\aSA$ not only depends on the quadratic scaling of the running time with the problem size, $t=AN^2$, but also on the prefactor $A$, as can be appreciated in Fig.~\ref{fig:SA_transition_other_scalings}, where data for $A=2$ and $A=8$ are reported.
Let us call $\aSA^{N^2}(A)$ the algorithmic threshold obtained in the large $N$ limit, scaling running times as $t=A N^2$.
We expect a range of thresholds for SA running in quadratic times, $\aSA^{N^2} \in [\aSA^{N^2}(0),\aSA^{N^2}(\infty)]$.
In particular, the lower value $\aSA^{N^2}(0)$ is the algorithmic threshold for SA running in linear time, which is hard to compute studying the linear regime, as it needs extrapolations, but is computable coming from the quadratic regime.
The upper value $\aSA^{N^2}(\infty)$ does correspond to the algorithmic threshold for SA running in quadratic time. Although it is in practice unreachable, we can get a bound from the best results in Fig.~\ref{fig:SA_transition_other_scalings}. The scaling $t=8N^{2}$ provides $\aSA^{N^2}(\infty) \gtrsim 4.045$. This critical value has been computed from the crossings of $\Uf$ curves, which are, in general, more accurate and robust with respect to $\Psat$ data.
This value is just a lower bound to the best algorithmic threshold for SA running in a quadratic time. However, we have noticed that $\aSA^{N^{2}}$ does not change much by varying the prefactor $A$.

\begin{figure}[t]
\centering
\includegraphics[width=0.4\textwidth]{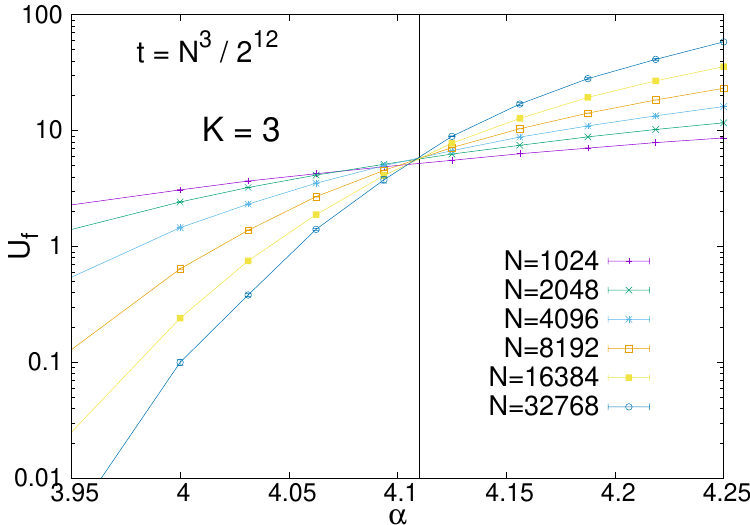}
\includegraphics[width=0.4\textwidth]{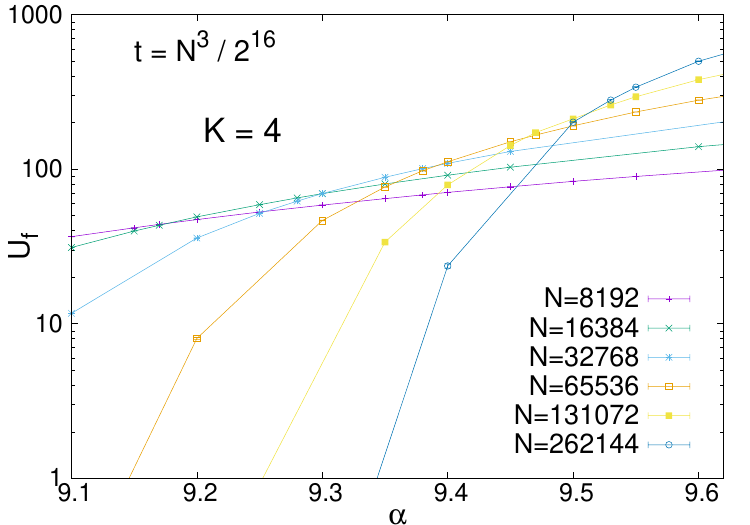}
\caption{{\bf Simulated annealing thresholds for cubic time scaling in random 3-Sat and random 4-Sat}. For each system size, we measure the average extensive energy $\Uf$ after a running time $t$ scaling cubically with the problem size. For $K=3$ (top panel) we used $t=N^{3}/2^{12}$, while for $K=4$ (bottom panel) we have $t=N^{3}/2^{16}$. The predicted algorithmic threshold for $K=3$ is located at the crossing points of the curves in the upper panel, $\aSA^{N^{3}} = 4.110(4)$.  In random 4-Sat, finite size effects are strong and estimating the algorithmic threshold requires a more accurate analysis and extrapolation (see text and Fig.~\ref{fig:fits_K_4}).}
\label{fig:SA_transition_other_scalings_cubic}
\end{figure}

A way of getting an upper bound to $\aSA^{N^{2}}$ is to study the performance of SA running in cubic times, $t=O(N^3)$.
The results are reported in Fig.~\ref{fig:SA_transition_other_scalings_cubic} for random 3-Sat and random 4-Sat. Data in the upper panel shows a clear crossing, implying $\aSA^{N^3} \simeq 4.11$ (the threshold depends again on the prefactor, but such a dependence is very mild, and we may ignore it for our purposes). Interestingly enough, this threshold is far beyond the dynamical transition $\alpha_d =3.86$ \cite{Montanari_2008}, and very close to what can be achieved by the best local search algorithm, Focused Metropolis Search (FMS), discussed in Appendix~\ref{app:FMS}.

Data in the lower panel of Fig.~\ref{fig:SA_transition_other_scalings_cubic} are for SA solving random 4-Sat in cubic time and present strong finite-size corrections that manifest themselves by the large shifting in the crossing points $\acr(N, 2N)$ between data of size $N$ and $2N$.
The reason for these large finite-size corrections can be understood by looking at the residual energy $\Uf$ at the crossing points. In the random 3-Sat case, the residual energy at the (unique) crossing point is small, $\Uf=O(1)$ (remember we are considering the extensive energy). Instead, for random 4-sat, $\Uf$ at the crossing points grows to much larger values. This, in turn, requires the study of larger systems to actually access the relevant values of $\Uf$ at criticality. Hence, the severe finite-size effects.

\begin{figure}[t]
\centering
\includegraphics[width=0.8\columnwidth]{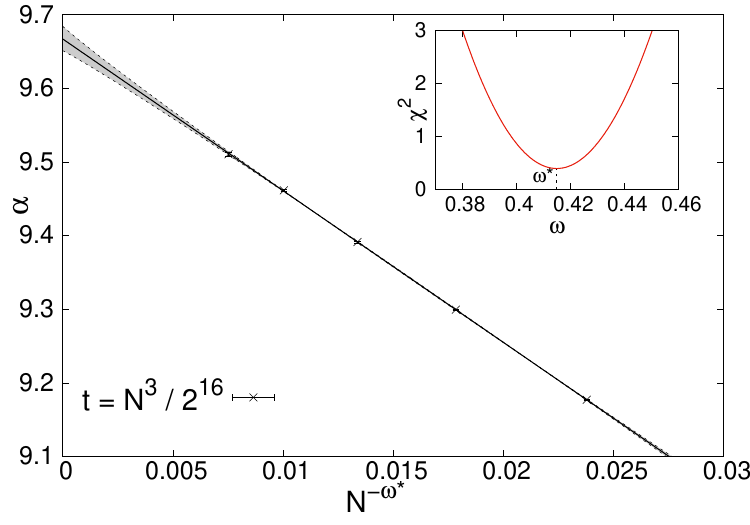}
\includegraphics[width=0.8\columnwidth]{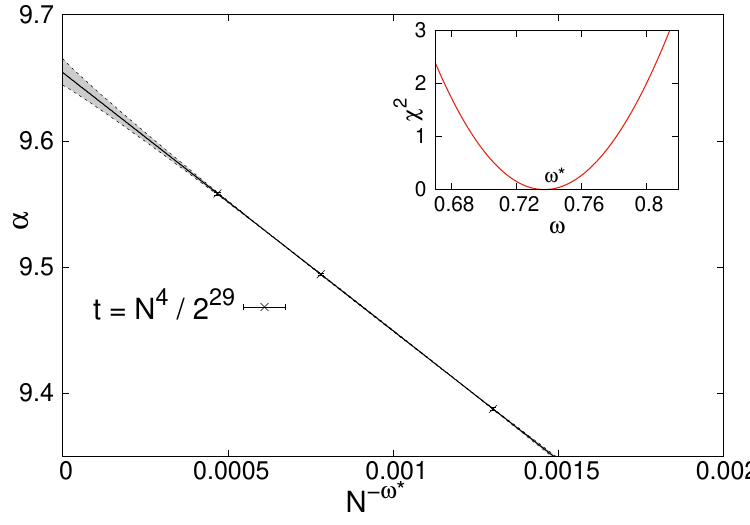}
\caption{{\bf Determination of the algorithmic thresholds for different time scalings in the random 4-Sat}. We include cubic (top panel) and quartic (bottom panel) scalings. Each point in the panels corresponds to the crossing $\acr(N, 2N)$ between the curves $\Uf(\alpha)$ for sizes $N$ and $2N$ (for cubic scaling are shown in the bottom panel of Fig.~\ref{fig:SA_transition_other_scalings_cubic}, for quartic scaling are not shown). Continuous lines are linear fits of the type $\acr(N, 2N)=\aSA-c N^{-\omega}$ made with the optimal exponent $\omega^{\ast}$ that minimizes the quality of the fit $\chi^2(\omega)$, shown in the inset. The shaded regions show the uncertainty on the fit considering statistically acceptable values of $\omega$ around $\omega^{\ast}$ (see the text for more details). Final results are $\omega^{\ast}=0.4149$ and $\aSA^{N^{3}}=9.665(15)$ for cubic scaling and  $\omega^{\ast}=0.7373$ and $\aSA^{N^{4}}=9.650(15)$ for quartic scaling.}
\label{fig:fits_K_4}
\end{figure}

A quantitative estimate of the algorithmic thresholds for $K=4$ requires the extrapolation of the crossing points $\acr(N, 2N)$ in the large $N$ limit. In the upper panel of Fig.~\ref{fig:fits_K_4} the crossing points are plotted as a function of the correction to scaling term $N^{-\omega}$. The optimal value for $\omega$ has been chosen by minimizing the $\chi^2$ value of the linear fit to $\acr(N, 2N)$ versus $N^{-\omega}$ data (as shown in the inset). 
There are two sources of error in this extrapolation.
The crossing values have a statistical uncertainty associated with the measurement of the energies $\Uf$ (the corresponding error bars are shown in Fig.~\ref{fig:fits_K_4}) but these are very small.
The largest uncertainty in the extrapolation comes from the unknown value of the correction-to-scaling exponent $\omega$.
For each value of $\omega$ we have performed a linear fit to the data (as shown in Fig.~\ref{fig:fits_K_4}) and computed the quality of the fit through the $\chi^2$ (shown in the inset).
The optimal value $\omega^{\ast}$ is the minimizer of $\chi^2(\omega)$.
However, all values of $\omega$ such that $\chi^{2}(\omega)-\chi^{2}(\omega^{\ast}) < 1$ are statistically acceptable. The colored regions in Fig.~\ref{fig:fits_K_4} represent the fits obtained using these acceptable $\omega$ values and provide the correct uncertainty on the extrapolated value.
This procedure provides the algorithmic thresholds $\aSA^{N^{3}}=9.665(15)$ and  $\aSA^{N^{4}}=9.650(15)$ for cubic and quartic scaling respectively.
The observation that $\aSA^{N^{3}}$ and $\aSA^{N^{4}}$ are compatible within errors implies that moving from the cubic to the quartic scaling, there is no improvement. We will come back later to discuss this point.

\begin{figure}[t]
\centering
\includegraphics[width=\columnwidth]{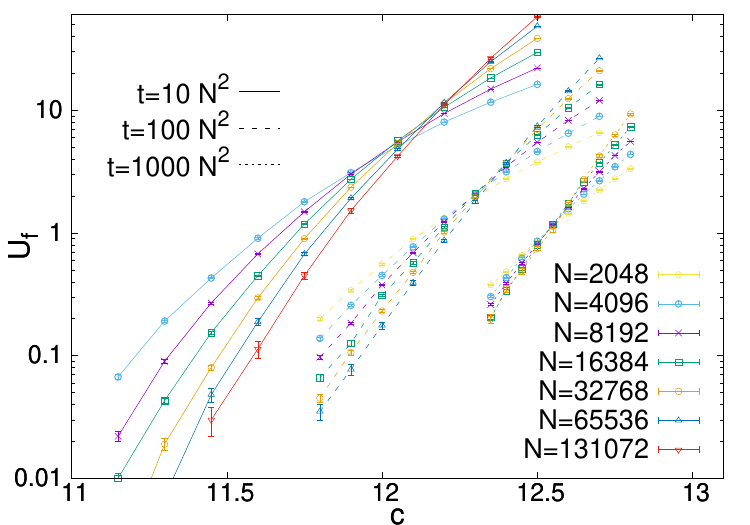}
\caption{{\bf Simulated annealing thresholds for quadratic time scaling in 5-coloring random graphs}. For each graph of size $N$ and mean degree $c$, we measure the average extensive energy $\Uf$ after a fixed running time, $t = A \, N^{2}$, with $A=10,10^2,10^3$ (from left to right). The crossing point identifies the SA algorithmic threshold.}
\label{fig:SA_transition_5col_other_scalings}
\end{figure}

To support the robustness of our findings, we consider a different combinatorial optimization problem, namely $q$-coloring of random graphs.
Given an Erd\"os-R\'enyi random graph of $N$ nodes and mean degree $c$, the problem is to assign to each node a color taking values in $\{1,2,...,q\}$, such as to avoid monochromatic edges connecting two nodes with the same color. The SA can also be applied to this problem, where the energy function counts the number of monochromatic edges in a given configuration of the variables.

Results for $q=5$ are reported in Fig.~\ref{fig:SA_transition_5col_other_scalings}, where we plot the final energy $\Uf$ reached by SA in a quadratic time, $t=AN^2$, as a function of graph mean degree $c$. As in the $K$-Sat case, we observe a clear crossing at the algorithmic threshold $\aSA$ (which again depends on the prefactor $A$). This confirms what we already obtained from $K$-Sat as a more general behavior.

\section{Algorithmic thresholds in the infinite-size limit}
\label{sec:tinfty}

\begin{figure*}[t]
\centering
\includegraphics[width=0.33\textwidth]{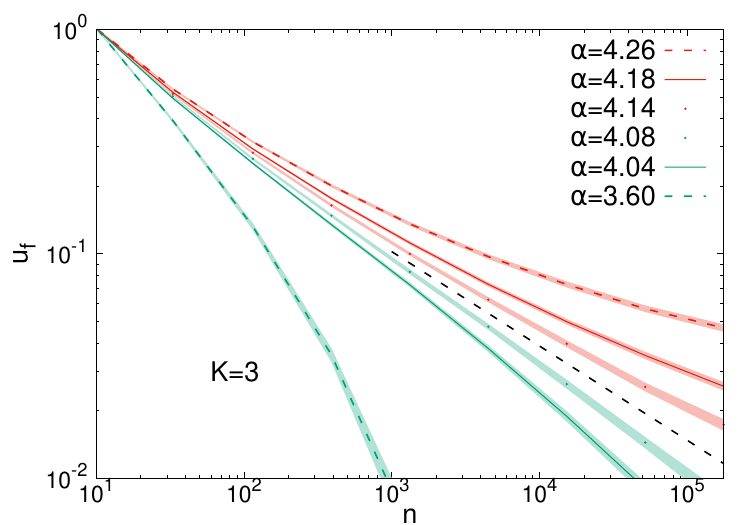}%
\includegraphics[width=0.33\textwidth]{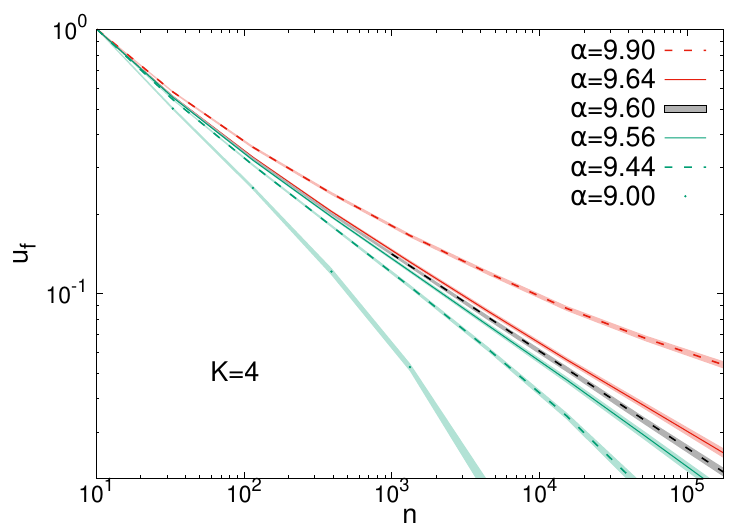}%
\includegraphics[width=0.33\textwidth]{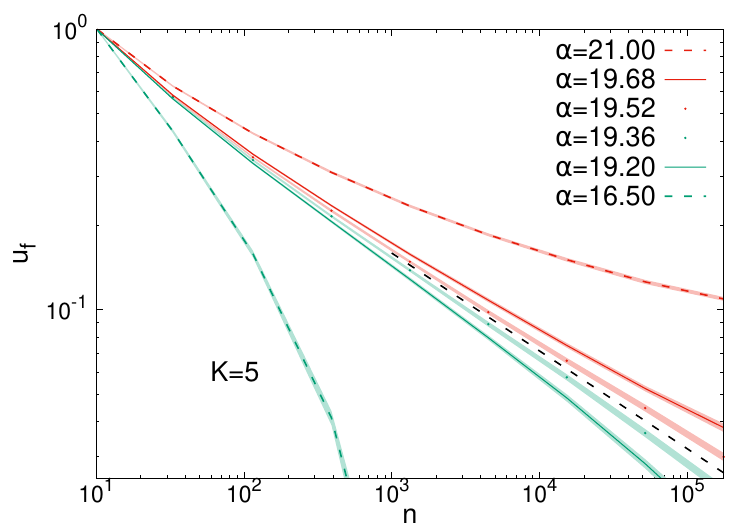}
\caption{\textbf{Estimating the algorithmic threshold in the infinite-size limit.} Simulated annealing with a linear temperature schedule is run for $n$ MCS steps to solve very large ($N=10^5$) random $K$-Sat instances with $K=3$ (left), $K=4$ (center) and $K=5$ (right). We report the final intensive energy $\uf$ (fraction of clauses violated) as a function of the total number $n$ of MCS. Axes are on a logarithmic scale to highlight when the function $\uf(n)$ follows a power-law decay at the critical algorithmic threshold. Green curves decay faster than a power law. Red curves decay slower than a power law. Shaded areas represent the measurement uncertainty (one standard deviation). The curve for $\alpha=9.60$ in the central panel is colored in black because it is not possible to distinguish it from a straight line when $n$ is large. Black dashed lines represent the critical power laws $u_f \propto n^{-b(K)}$ with $b(3)=-0.42$, $b(4)=-0.37$, and $b(5)=-0.35$. Estimated algorithmic thresholds are $\aSA^{\infty}(K=3) = 4.11(1)$, $\aSA^{\infty}(K=4) = 9.60(2)$ and $\aSA^{\infty}(K=5) = 19.44(3)$.}
\label{fig:transitions_fig}
\end{figure*}

Here, we discuss a complementary estimate of the algorithmic thresholds for SA in solving random $K$-Sat problems obtained in the so-called infinite-size limit. In practice, we simulate very large instances of size $N=10^5$ for $K=3,4,5$, such that finite size effects are negligible. We run SA as explained above: a linear schedule in temperature with $T_0=1$ and $T_n=0$ with 1 MCS per temperature (the most adiabatic scheduling), for a total of $n$ MCS. We measure the final energy $\Uf$, or equivalently its intensive value $\uf = \Uf / N$, as a function of the running time $n$. The results are presented in a doubly logarithmic scale in Fig.~\ref{fig:transitions_fig}.

For each value of $K=3,4,5$, we have run simulations in a range of $\alpha$ values which is fully contained in the SAT phase (the reader can find the values of the Sat-Unsat critical threshold $\alpha_s$ in Table \ref{tb:table_alphas}). This means that the equilibrium energy is exactly zero. However, when the SA is run for a finite number $n$ of MCS in a very large problem (remember we are working in the infinite size limit), it will soon or later fall out of equilibrium at low temperatures, and the final energy $\Uf(n)$ will be strictly positive.

\begin{figure}[t]
\centering
\includegraphics[width=0.4\textwidth]{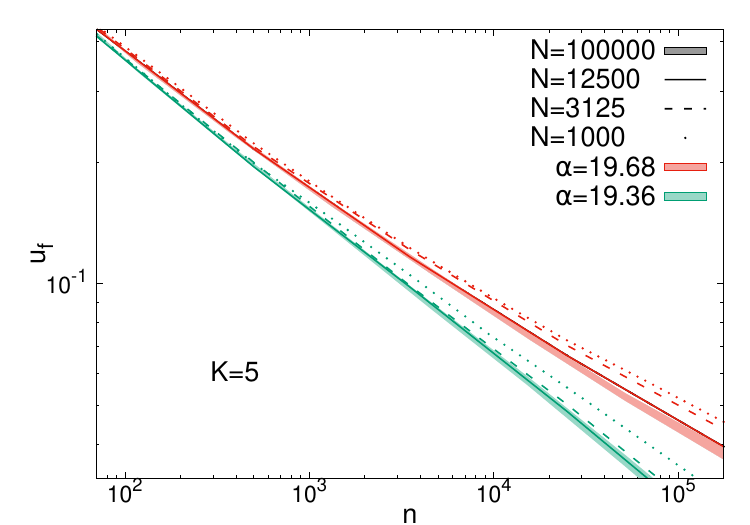}
\caption{\textbf{Finite size effects on simulated annealing behavior.} The infinite-size decay measured on problems of size $N=10^5$ persists also for smaller sizes, $N\sim O(10^4)$. On even smaller sizes, the slope changes a little, but the curvature is preserved, thus making the proposed procedure to estimate $\aSA^\infty$ robust. Error bars are reported only for the largest size with a shadowed area.}
\label{fig:FSE_fig}
\end{figure}

In the large size limit, the intensive energy $\uf(n)$ is a decreasing function of $n$ with practically no fluctuations at any finite value of $n$ (this is a consequence of the energy being a self-averaging quantity). Indeed, curves in Fig.~\ref{fig:transitions_fig}, which have been measured for problems of size $N=10^5$, are smooth and very close to the infinite-size limiting curve $\uf(n)$ (finite size effects are discussed below when commenting Fig.~\ref{fig:FSE_fig}).

In Fig.~\ref{fig:transitions_fig}, we plot the data in a doubly logarithmic scale because we want to stress which curves decay faster than a power law (green curves) and which ones decay slower than a power law (red curves).
We observe that in a small range of $\alpha$ the curvature of the data shown in Fig.~\ref{fig:transitions_fig} changes from negative (green curves) to positive (red curves).
This observation suggests that the energy decays as a power law $\uf(n) \propto n^{-b}$ at a single value of $\alpha$ (or in a very narrow interval of $\alpha$ values). Let us call critical such $\alpha$ value (or interval of).

We argue that the algorithmic threshold $\aSA^\infty$ corresponds to the critical $\alpha$ value (or to the upper limit of the critical interval).
Our argument goes as follows: first, we prove that at criticality the time to find a solution by SA in a problem of size $N$ scales like a power law, $t \sim N^{1+1/b}$; then, we prove that for $\alpha>\aSA^\infty$ the time to find a solution scales super-polynomially.

At a critical $\alpha$ value, we have $\uf(n) \propto n^{-b}$ in the infinite size limit.
Thus, the typical time to reach a solution in a problem of size $N$, which is finite, but large enough, is given by the $n$ value such that $\uf(n) \sim 1/N$. Indeed, $\uf = a/N$ implies $\Uf=a$, and if $a<1$ then there is a non-zero probability that SA finds a solution (to have a mean extensive energy equal to $\Uf=a<1$, at least a fraction $1-a$ of samples must have zero energy, given the energy takes only integer values).
This argument states that SA finds solutions with a non-zero probability if run with $n \sim N^{1/b}$ MCS, which requires a time $t \sim N^z$ with a dynamical exponent $z=1+1/b$, considering that every MCS takes a time $O(N)$.

For $\alpha>\aSA^\infty$ the curve $\uf(n)$ has a positive curvature (in a log-log scale), which means that the local slope decreases as $n$ grows. We can use again the argument above, but now the effective $b$ exponent becomes smaller for larger $n$, and eventually goes to zero for $n\to\infty$. This implies that the typical time to reach a solution grows faster than any power law, thus proving SA is a super-polynomial algorithm in this regime.

We estimate the algorithmic thresholds for SA in the infinite-size limit $\aSA^\infty(K)$ by bounding the critical $\alpha$ values.
In Fig.~\ref{fig:transitions_fig} we report data for $\uf(n)$ measured with $K=3,4,5$ (panels from left to right). In each case, the critical $\alpha$ value is between the higher green curve and the lower red one. The estimated algorithmic thresholds are $\aSA^{\infty}(3) = 4.11(1)$, $\aSA^{\infty}(4) = 9.60(2)$ and $\aSA^{\infty}(5) = 19.44(3)$.

From the power law decay at criticality, we can estimate the dynamical exponent $z(K)$ ruling the scaling of times with problem size, $t \propto N^z$.
Our estimates are $z(3)=3.4(3)$, $z(4)=3.7(2)$ and $z(5)=3.9(1)$.
We notice that all exponents are between 3 and 4 (and this is an argument supporting the lack of any gain in going from the cubic to the quartic scaling, observed in Fig.~\ref{fig:fits_K_4} and discussed in more detail below).

The procedure we have proposed and used to estimate $\aSA^\infty$ in the infinite-size limit assumes finite-size effects are somehow benign. We check our assumption in Fig.~\ref{fig:FSE_fig} through the study of the decay of $\uf(n)$ in the large $n$ regime for very different problem sizes $N$. The largest size reported, $N=10^5$, is the same one used in Fig.~\ref{fig:transitions_fig} to estimate $\aSA^\infty$ and comes with a shaded area representing the statistical uncertainty (one standard deviation).
The first important observation is that the data collected in solving problems of a size smaller by an order of magnitude, $N\sim O(10^4)$, still follow closely the infinite-size decay. 
The second observation is that, for even smaller problems, although finite-size effects are visible, the curvature (in a log-log scale) is preserved, i.e.\ it is negative for $\alpha < \aSA^\infty$ and positive for $\alpha > \aSA^\infty$.
These observations imply our procedure to estimate $\aSA^\infty$ is robust and can be applied even on problems of a modest size.

\section{Summary and discussion}

\begin{table}[t]
\begin{tabular}{|c|c|c|c|c|c|c|}
\hline
$K$ & $\alpha_d$ & $\alpha_c$ & $\alpha_s$ & $\aSA^{N^2}$ & $\aSA^{N^3}$ & $\aSA^{\infty}$ \\
\hline
3  & 3.86  & 3.86 & 4.267 & 4.025 -- 4.045 & 4.110(4) & 4.11(1) \\
\hline
4  & 9.38 & 9.547 & 9.931 & 9.15 -- 9.4 & 9.665(15) & 9.60(2) \\
\hline
5 & 19.16 & 20.80 & 21.117 & --- &  --- & 19.44(3) \\
\hline
\end{tabular}
\caption{\textbf{Summary of thresholds for random $K$-Sat.} The values for the dynamical transition $\alpha_d$ and the condensation transition $\alpha_c$ have been taken from Table 1 in Ref.~\cite{Montanari_2008}. The Sat-Unsat thresholds $\alpha_s$ are from Ref.~\cite{MertensKSAT2006}. We report the algorithmic thresholds for SA, obtained from finite-size scaling in the quadratic regime, $\aSA^{N^2}$, with a prefactor $A\in[0.25,8]$, from finite-size scaling in the cubic regime, $\aSA^{N^3}$, and in the infinite-size limit, $\aSA^\infty$.}
\label{tb:table_alphas}
\end{table}

We start summarizing our results. The numerical values of the algorithmic thresholds we have estimated are reported in Table~\ref{tb:table_alphas}, together with other relevant thresholds obtained from the thermodynamic solution to the random $K$-Sat problem \cite{MertensKSAT2006,KrzakalaKSAT2007,Montanari_2008} and described in the Introduction.

The algorithmic thresholds for SA in solving random $K$-Sat problems can be well defined only for super-linear regimes (we discuss in App.~\ref{app:linearRegime} the linear regime).
The estimates for $\aSA$ in the quadratic and the cubic regimes are clearly different, as can be seen from the thresholds summarized in Table~\ref{tb:table_alphas}.
This key result requires a revision of the concept of algorithmic threshold, which can no longer be considered unique, given the problem and the algorithm, but it can depend on the scaling between the problem size and the resolution time.

The algorithmic threshold grows when moving from a quadratic to a cubic scaling regime. However, as shown in Fig.~\ref{fig:fits_K_4}, going to a quartic scaling regime is not helpful.
A possible explanation for this observation comes from the estimates we got via the analysis of the infinite-size limit.
Let us consider that the algorithmic threshold $\aSA^\infty$ obtained from the latter provides the largest threshold for any polynomial scaling regime, that is, for $\alpha>\aSA^\infty$ the SA running times grow super-polynomially. Given that at $\aSA^\infty$ the SA running times scale like $t \propto N^z$ with $z<4$ (for the $K$ values studied), then running SA for a time growing quartically with $N$ can not push the algorithmic threshold beyond $\aSA^\infty$. Moreover, since $\aSA^\infty$ is statistically compatible with $\aSA^{N^3}$, we conclude that thresholds in the cubic and quartic regimes should be statistically compatible.

Among all possible polynomial time regimes, we observe the cubic one to be optimal. This is again related to the exponent $z$ measured at criticality in the infinite-size limit, which is slightly larger than 3. The optimal $\aSA$ thresholds can be compared with the phase transitions in the structure of solutions.
For every value of $K$ studied, the optimal $\aSA$ is located well above the dynamical transition $\alpha_d$.
This is a non-trivial observation. Indeed, given that at $\alpha_d$ the ergodicity of MC-based algorithms is broken, one could imagine $\alpha_d$ to be the algorithmic threshold for SA and any MC-based algorithm \cite{mezardksat2002,KrzakalaKSAT2007}.
However, the ergodicity breaking happening at $\alpha_d$ is related to the sampling of the equilibrium measure, which is uniformly distributed over all the solutions. By sampling solutions according to a non-uniform measure, it is well known that the dynamical threshold can change \cite{Budzynski_2019,cavaliere2021optimization}. The observation that SA can efficiently solve random $K$-Sat instances well beyond $\alpha_d$ implies SA is not sampling solutions uniformly, but in a more efficient way that accesses a subset of solutions even beyond $\alpha_d$.

Although we have chosen to study the SA algorithm for its generality, rather than for its performance, it is still worth comparing it to the best local search algorithm, FMS, discussed in App.~\ref{app:FMS}. FMS thresholds in the linear regime are $\aFMS \simeq 4.2$ for $K=3$ and $\aFMS \simeq 9.62$ for $K=4$, and grow to $\aFMS \simeq 4.22$ for $K=3$ and $\aFMS \simeq 9.7$ for $K=4$ in a quadratic regime. Compared to the FMS thresholds, the performance of SA are suboptimal (as expected) but not too far from optimality.
Thus, restricting our study to the SA algorithm has not been a too detrimental choice in terms of performance.

We discuss now two fundamental questions. Why is the optimal performance of SA (and probably of many other algorithms) achieved in a super-linear regime, with running times scaling with a non-trivial power of the problem size, $ t\sim N^z$? How can several critical regimes with different time scaling (e.g., quadratic and cubic) coexist?
These questions are non-trivial because, in the simplest scenario, one could imagine the existence of just two phases separated by an algorithmic threshold $\aalg$: an ``easy'' phase for $\alpha<\aalg$ with no algorithmic barriers, where the algorithm converges to a solution quickly\footnote{How quickly may depend on the nature of the algorithm, but most often in a time $O(N)$ or $O(N \log(N))$.}, and a ``hard'' phase for $\alpha>\aalg$ where the presence of algorithmic barriers impedes fast convergence to a solution an requires super-polynomial times (e.g.\ when an extensive free-energy barrier needs to be surpassed by a stochastic algorithm satisfying detailed balance). In this simple scenario, the algorithmic threshold $\aalg$ is unique for any polynomial scaling.

The results presented in this work reveal a novel scenario, where the easy phase is itself divided into different regions that are accessible only by scaling running times with the proper power law of the problem size. Roughly speaking, we can say that the easy phase gets broken into easier and less easy phases. Every transition between two of these polynomially accessible phases corresponds to a different algorithmic threshold between polynomial regimes, while the largest among these thresholds corresponds to the ultimate transition to a super-polynomial regime.

We conjecture that some (if not all) of these polynomially accessible phases are controlled by entropic barriers. Indeed, as shown in App.~\ref{app:linearRegime}, there are greedy algorithms that are not allowed to climb over energetic barriers and still find solutions in the region accessible by SA, scaling times quadratically with the problem size. This observation strongly suggests that, in this `quadratic' region, solutions are difficult to access by SA only because the stochastic dynamic has to find the right path in a complex energy landscape with many apparently equivalent directions (i.e., an entropic barrier).

Increasing $\alpha$, the energy landscape may become more and more complicated to explore for purely entropic reasons, even without developing relevant energetic barriers. So, a plausible explanation for the many different algorithmic thresholds we have found is that, even when some solutions are polynomially accessible, these are few and connected by tiny ``corridors'' to the rest of the configurational space, so as to make it very hard to find them, especially by a purely stochastic algorithm, like SA (we discuss in App.~\ref{app:FMS} that a focused algorithm performs much better, probably because it can enter in these tiny and rare corridors more easily).

The picture we have in mind for the space of solutions in these polynomially accessible phases above $\alpha_d$ is similar to the one found in other models, where it has been proved that regions exist such that the majority of solutions are clustered and can not be sampled efficiently, but a subdominant subset of solutions can be algorithmically achieved. This is the case, for example, of the binary perceptron, where the majority of solutions are isolated and unaccessible \cite{krauth1989storage,huang2014origin}, but algorithms can still work \cite{braunstein2006learning} by reaching the few accessible states, which are ``flat'' minima of very large local entropy connected by paths of low probability \cite{krauth1989storage, braunstein2006learning, huang2014origin, baldassi2015subdominant, baldassi2016unreasonable, baldassi2021unveiling, baldassi2023typical}.

A scenario in which the accessible states that can be algorithmically reached are flat, i.e., marginal, is very common in disordered systems \cite{folena2020rethinking}. Assuming such a scenario, the time scale needed to solve a problem by SA (or by any other algorithm) is likely to be related to the entropic difficulty of finding a descending path in a very flat landscape, rather than to the hardness of jumping over an energetic barrier. In such a scenario, the dynamical exponent $z$ ruling the scaling between problem size and time to find a solution is probably related to the dimension of the marginal manifold where the searching dynamics is taking place and the dimension (i.e., the number) of descending directions that eventually lead to a solution. Future studies will better clarify this relation.
However, it should be noted that SA is often out of equilibrium and does not sample solutions according to the uniform measure \cite{Budzynski_2019}. Thus, analytic computations are much harder than in equilibrium statistical physics.

We conclude by remarking that the methodology presented in this work can be easily applied to any other combinatorial optimization problem.
We hope that it will become the standard procedure for computing the algorithmic thresholds in several problems solved by different algorithms.

Among the problems to which our procedure can be extended, it is worth pointing out artificial neural networks. Indeed, in this class of problems, both the training time and the inference time can be scaled with the problem size to achieve optimal performances \cite{snell2024scaling, brown2024largelanguagemonkeysscaling}, but to the best of our knowledge a systematic study of different time-scaling regimes is lacking.

Another interesting case are inference problems, that can be recast as optimization problems with a known optimum (corresponding to the signal to be inferred). Recent findings in the literature rigorously proved that, in some cases, superlinear algorithms can do better than linear ones, the latter also including message-passing algorithms whose running times typically scale like $O(N \log(N))$ \cite{zadik2022lattice, song2021cryptographic, angelini2018parallel, wein2019kikuchi}.
The results in the present work strongly suggest that the optimal algorithmic performance requires higher-order time-scaling (e.g., quadratic or cubic in $N$) in random optimization problems. It would be very interesting to extend the present study to the determination of thresholds for MC algorithms run at different time-scales in solving inference problems.

We hope that our work can help in building a new general framework to explain these results. We point out that the recent OGP theory for algorithmic hardness \cite{gamarnik2021overlap} is not enough, because MC algorithms running in super-linear time are not stable algorithms, and thus their performances are not ruled by the OGP.

\begin{acknowledgments}
This research has received financial support from the ``National Centre for HPC, Big Data and Quantum Computing - HPC'', Project CN\_00000013, CUP B83C22002940006, NRP Mission 4 Component 2 Investment 1.5,  Funded by the European Union - NextGenerationEU and by PRIN 2022 PNRR, Project P20229PBZR, CUP
B53D23028410001, Mission 4, Component C2, Investment 1.1,  Funded by the European Union - NextGenerationEU. The numerical simulations were conducted using the DARIAH HPC cluster at CNR-NANOTEC in Lecce, funded by the "MUR PON Ricerca e Innovazione 2014-2020" project, code PIR01\_00022.
\end{acknowledgments}

\bibliography{ref_cvme_EPL}

\appendix

\begin{figure}[ht]
\centering
\includegraphics[width=0.4\textwidth]{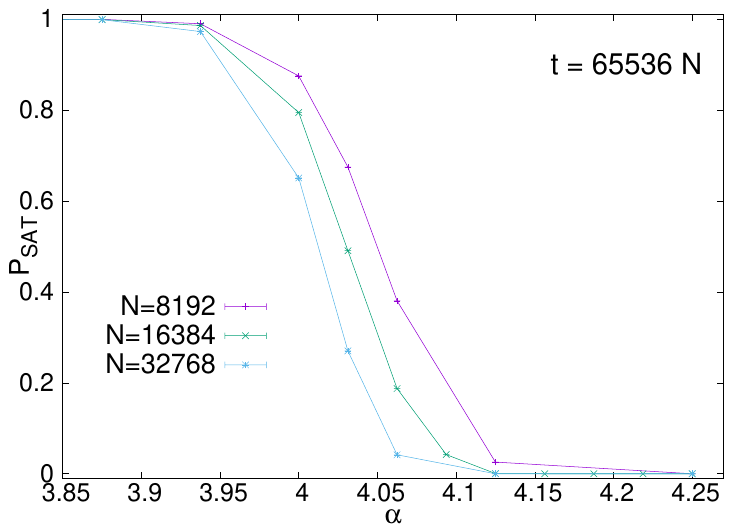}
\includegraphics[width=0.4\textwidth]{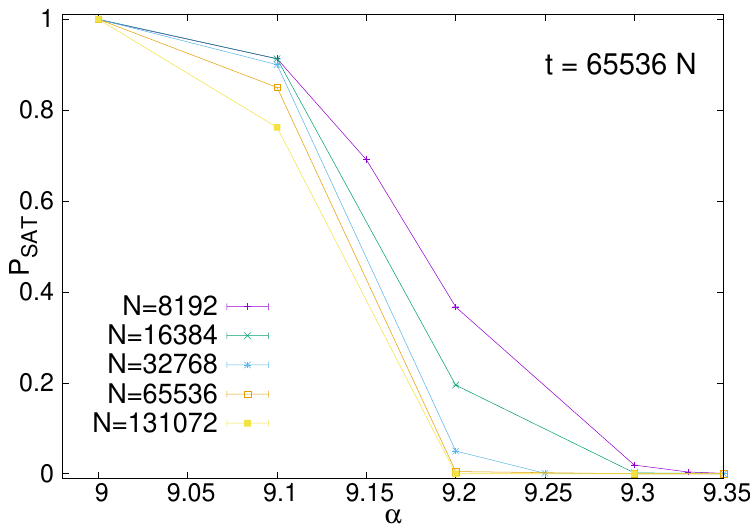}
\caption{{\bf Performances of Simulated Annealing in solving random K-Sat problems in linear time.} The probability that SA finds a solution in a fixed ($2^{16}$) number of MCS  is plotted as a function of $\alpha$ for $K=3$ (upper panel) and $K=4$ (lower panel). The curves move to the left by increasing system size and have no crossing at all, thus making it hard to estimate the corresponding algorithmic threshold.}
\label{fig:PSAT_linear}
\end{figure}

\section{Estimating algorithmic thresholds in the (quasi-)linear time regime}
\label{app:linearRegime}

Here we first provide evidence that studying the performance of SA in the linear time regime, i.e., when SA is run for a \emph{fixed} number of MCS, is complicated and does not provide a clear estimate for the algorithmic threshold. Then we show that a \emph{greedy} version of the MC sampling, where the temperature is set to zero and the algorithm can never increase the energy, can still find solutions in almost linear time.

In Fig.~\ref{fig:PSAT_linear} we plot the probability that SA finds a solution after being run for a fixed number of MCS, that is, for a time growing linearly with the problem size $N$. We notice that both for $K=3$ (upper panel) and $K=4$ (lower panel) the curves move leftwards when the problem size is increased. The explanation for this observation is straightforward: the same number of MCS may be enough to solve a smaller problem, but may not suffice to solve a larger problem. As a consequence, the data plotted in Fig.~\ref{fig:PSAT_linear} show no crossing at all and just provide \emph{upper bounds} to the actual algorithmic threshold in the linear regime. A more precise estimate for such a threshold would require uncontrolled extrapolations (there is no theory for these finite size corrections) and would eventually provide an estimate with a large systematic error.

This is the reason why we started exploring super-linear regimes in SA solving random $K$-Sat and we discovered that clear algorithmic thresholds for SA can be obtained only in these super-linear regimes (and this is the main point of our work).

\subsection{Performances of a quasi-linear greedy algorithm}

Let us consider now the limit $T\to0$, or equivalently $\beta\to\infty$, in Eq.~\eqref{eq:Metropolis_rates}. The resulting Metropolis algorithm will evolve the configuration without ever increasing the system energy. This greedy algorithm is known as zero-temperature Monte Carlo (ZTMC), and it also has a special role in the interplay between statistical physics and combinatorial optimization \cite{MezardParisiVirasoro, mezardksat2002, mezard2009information}.

\begin{figure}[t]
\centering
\includegraphics[width=0.4\textwidth]{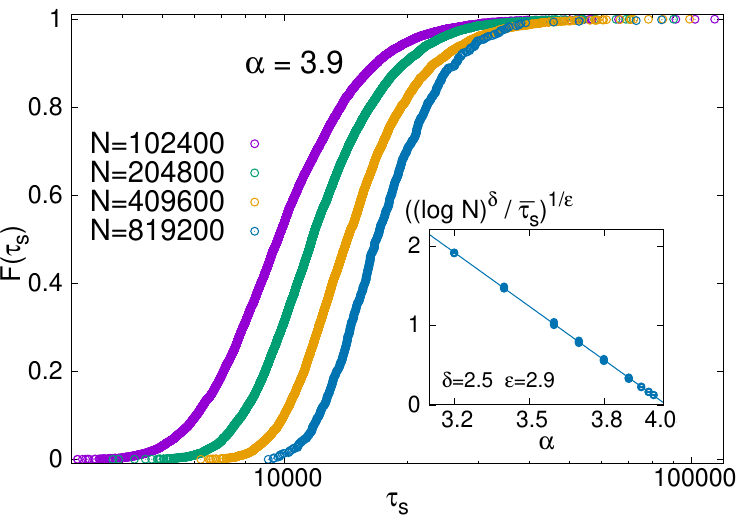}
\includegraphics[width=0.4\textwidth]{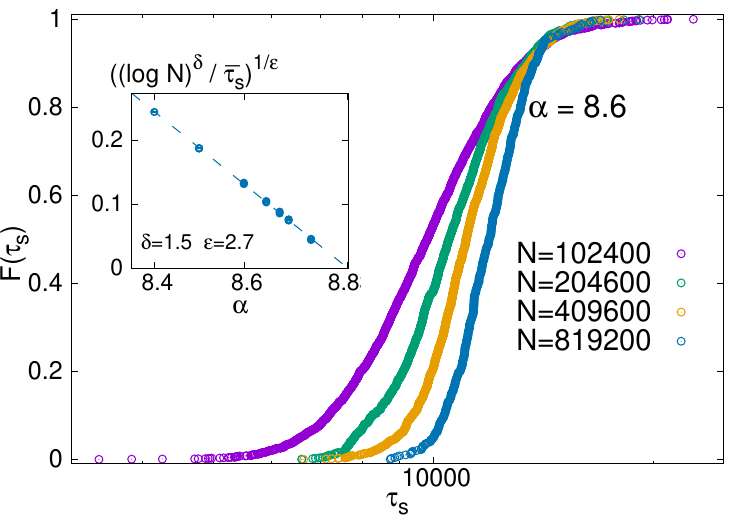}
\caption{{\bf Algorithmic threshold for the greedy zero-temperature Monte Carlo in solving random $K$-Sat problems} with $K=3$ (top panel) and $K=4$ (bottom panel). The main panels show the cumulative distribution of the time $\tau_s$ needed to reach a solution. In both cases, the $\alpha$ value has been chosen to be a few percent below the algorithmic threshold. By increasing the problem size, at a fixed $\alpha$ value, the mean time $\overline{\tau_s}$ grows like $(\ln N)^\delta$.  In the insets, we report the fits estimating the algorithmic threshold $\aZTMC$, assuming the mean value follows the law $\overline{\tau_s} \sim (\ln N)^{\delta} \, (\alpha - \aZTMC)^{-\varepsilon}$. Best fit parameters are $\aZTMC \simeq 4.04$ with $\delta=2.5$ and $\varepsilon=2.9$ for $K=3$, and $\aZTMC \simeq 8.83$ with $\delta=1.5$ and $\varepsilon=2.7$ for $K=4$.}
\label{fig:MC_T0_thresholds}
\end{figure}

For several values of $\alpha$ and $N$ with both $K=3$ and $K=4$, we have run the ZTMC algorithm until convergence to a solution (restring to the Sat phase soon or later this will happen). The results have been averaged over $10^4$ problems for each value of $\alpha$ and $N$, but for the largest size ($N=819200$), in which case, only 500 problems were studied.

In Fig. \ref{fig:MC_T0_thresholds}, we report some results obtained by running ZTMC at a reasonably large value of $\alpha$ (a few percent below the estimated algorithmic threshold, $\aZTMC$) for both $K=3$ (upper panel) and $K=4$ (lower panel). In the main panels and for a fixed value of $\alpha$, one can see that the cumulative distribution $F(\tau_s)$ of the times to reach a solution shifts to the right when the system size is increased. Both the median and the mean time to reach a solution scale with a small power of $\ln(N)$. Since the time unit we use is a Monte Carlo sweep, which already scales linearly with $N$, the average solution time $\overline{\tau_s}$ for the ZTMC grows slightly faster than linearly in $N$.

We observe that, also in this case, the estimation of the algorithmic threshold requires some extrapolation, given that the curves in Fig. \ref{fig:MC_T0_thresholds}  move right without any crossing.
So, to estimate the algorithmic threshold, we assume that the mean time to solution follows the law  $\overline{\tau_s} \sim (\ln N)^{\delta} \, (\alpha - \aZTMC)^{-\varepsilon}$ with fixed exponents $\delta$ and $\varepsilon$ for every $K$ value. In the insets of Fig.~\ref{fig:MC_T0_thresholds}, we plot $\overline{\tau_s}$ scaled by $(\ln N)^\delta$ such that, for every value of $\alpha$, data at different values of $N$ perfectly collapse. The collapsed data can be further scaled (by raising it to the power $1/\varepsilon$) such that they align well and can be safely interpolated by a linear fit. We estimate the algorithmic threshold for the ZTMC algorithm by the point where the linear fit extrapolates to zero. The best estimates are $\aZTMC \simeq 4.04$ using  $\delta=2.5$ and $\varepsilon=2.9$ for $K=3$, and $\aZTMC \simeq 8.83$ using $\delta=1.5$ and $\varepsilon=2.7$ for $K=4$.

Now, where does the $(\ln N)^{\delta}$ factor come from in an algorithm without any thermal noise? If the ZTMC were always taking steps to decrease the energy, it would not take more than linear time to find a local energy minimum. Indeed, the initial energy of a random configuration is $2^{-K}\alpha N$. If the ZTMC algorithm would always decrease the energy at every step, it would reach the energy minimum in linear time. However, the ZTMC algorithm can also move between configurations with the same energy. When the algorithm reaches a region with many flat directions, it may take some time before finding a good direction to decrease further the energy. We numerically find that this time scales like $(\ln N)^{\delta}$.
So, even with a greedy algorithm, it is not possible to find a solution in linear time because of the existence of entropic barriers (flat regions of the energy landscape).
 
Although the ZTMC algorithmic threshold for $K=4$ is below $\alpha_d$ and well below the best results of SA, it is surprising that for $K=3$ this greedy algorithm goes well beyond $\alpha_d$ and its threshold is comparable, within the error, with that for the SA algorithm run with quadratic time scaling. The different behaviour between $K=3$ and $K=4$ could be due to the different type of clustering transition in the two cases, continuous for $K=3$ and discontinuous for $K=4$ \cite{Montanari_2008}.

\section{Algorithmic thresholds of Focused Metropolis Search}
\label{app:FMS}

\begin{figure*}[t]
\centering
\hfill
\includegraphics[width=0.4\textwidth]{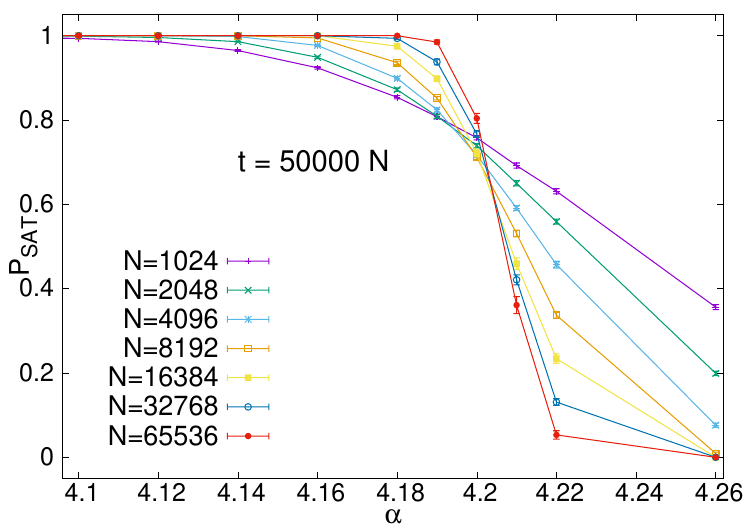}\hfill
\includegraphics[width=0.4\textwidth]{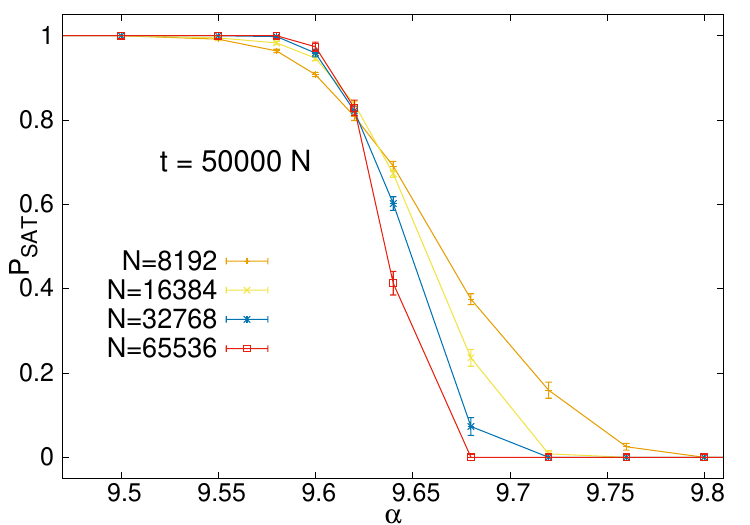}\hfill\phantom{.}\\
\hfill
\includegraphics[width=0.4\textwidth]{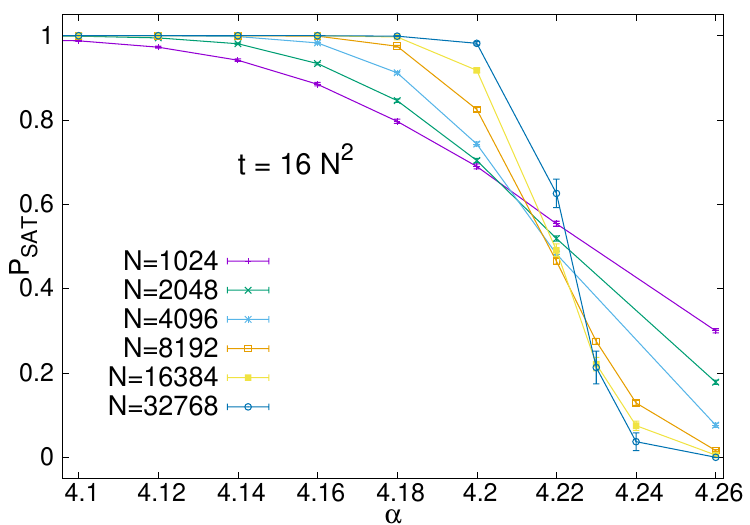}\hfill
\includegraphics[width=0.4\textwidth]{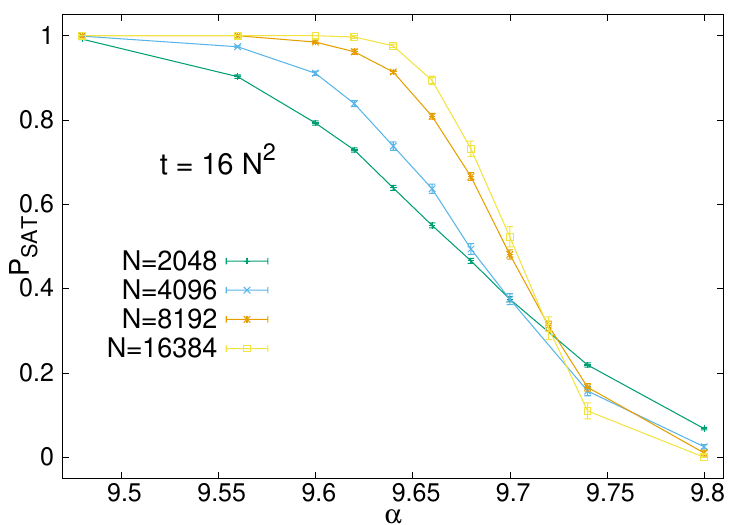}\hfill\phantom{.}
\caption{{\bf Focused Metropolis Search with linear and quadratic time scalings in random K-Sat}. For several system sizes, $N$, we plot the probability $\Psat$ to find a solution to random $K$-Sat instances ($K=3$ in left panels and $K=4$ in right panels) by running FMS for a time scaling linearly (upper panel) and quadratically (lower panel) with $N$. For the linear scaling, the algorithmic thresholds are $\aFMS^{N} \sim 4.2$ for $K=3$ and $\aFMS^{N} \sim 9.62$ for $K=4$.}
\label{fig:FMS}
\end{figure*}

Focused Metropolis Search (FMS) is a well-known efficient algorithm introduced in Ref. \cite{Seitz_2005} to solve $K$-Sat instances. It has been shown \cite{Seitz_2005} to perform on par with the best versions of the famous Walksat \cite{PapadimitrouWalkSAT, SelmanKautz1993, SKC1996}. 

Each iteration of FMS starts by selecting an unsatisfied clause in the Boolean formula, uniformly at random among all unsatisfied clauses. Then, the algorithm chooses a variable $x_i$ inside that clause also uniformly at random and computes the change $\Delta U$ in the energy that would be provoked by flipping $x_i$. If $\Delta U \leq 0$, the flip is always accepted; if $\Delta U > 0$ the flip is accepted with probability $\eta^{\Delta U}$. Here, $\eta \in [0, 1]$ is an algorithmic parameter interpreted as the level of noise in the dynamics. The reader should notice that, once the variable $x_i$ is selected, FMS follows the Metropolis rule with $\eta=e^{-\beta}$ (see Eq. (\ref{eq:Metropolis_rates})).

Remarkably, the numerical results presented in Refs.~\cite{Seitz_2005, AlavaPNAS2008} indicate that FMS is able to solve random $K$-Sat instances in linear time even beyond the condensation transition $\alpha_c$. In these works, the FMS algorithmic thresholds were estimated by locating the divergence of the time to solution. However, methods based on extrapolations may be quite noisy, and thus we estimate FMS thresholds using the more accurate method introduced in the main text. 

The upper panels of Fig.~\ref{fig:FMS} show the results for a linear time scaling. For different system sizes, $N$, we measured the probability that FMS finds a solution in no more than $5\cdot 10^4$ MCS. The curves have nice crossings close to $\aFMS^{N} \sim 4.2$ for $K=3$ and $\aFMS^{N}\sim 9.62$ for $K=4$. Please notice that FMS can solve random 3-Sat in linear time in almost all the Sat phase ($\alpha<\alpha_s=4.267$), outperforming SA with superlinear time scalings. For $K=4$, on the other hand, running FMS in linear time achieves a performance similar to that of running SA in superlinear time.

The quadratic scaling, $t=16 N^{2}$, shown in the bottom panels of Fig.~\ref{fig:FMS}, appreciably improves the performance obtained with a linear scaling. For $K=4$ and $N=16384$, FMS solves almost half of the instances with $\alpha=9.7$, an outstanding result for a local search algorithm in random K-Sat. Determining the corresponding algorithmic thresholds is more difficult in this case because of strong finite-size effects. 

The results in this appendix illustrate the generality of our method to estimate the algorithmic thresholds in hard combinatorial optimization problems. The more efficient FMS algorithm displays the same phenomenology discussed in the main text for the SA algorithm, with algorithmic thresholds depending on the time scaling.

\end{document}